# High-level synthesis under I/O Timing and Memory constraints

Philippe Coussy, Gwenole Corre, Pierre Bomel, Eric Senn, Eric Martin
LESTER LAB, UBS University, CNRS FRE 2734

*Abstract*—The design of complex Systems-on-Chips implies to take into account communication and memory access constraints for the integration of dedicated hardware accelerator. In this paper, we present a methodology and a tool that allow the High-Level Synthesis of DSP algorithm, under both I/O timing and memory constraints. Based on formal models and a generic architecture, this tool helps the designer to find a reasonable trade-off between both the required I/O timing behavior and the internal memory access parallelism of the circuit. The interest of our approach is demonstrated on the case study of a FFT algorithm.

## I. INTRODUCTION

Electronic design complexity has increased hugely since the birth of integrated circuits. System level technologies, over recent years, have moved from Application Specific Integrated Circuits (ASICs) and Application Specific Signal Processors (ASSPs) to complete System-On-Chip (SoC) designs. This increment in the chip complexity requires an equivalent shift in the design methodology and a more direct path from the functionality down to the silicon. In [1-3], the authors propose system synthesis approaches where the algorithms of the functional specification correspond to pre-designed components in a library. Macro generators produce the RTL architecture for hardware blocks by using the "generic"/ "generate" VHDL mechanisms: the synthesis process can hence be summarized as a block instantiation. However, though such components may be parameterizable, they rely on fixed architectural models with very restricted customization capabilities. This lack of flexibility in RTL blocks is especially true for both the communication unit, which I/O scheduling and/or I/O timing requirements are defined, and the memory unit, which data distribution is set.

High-Level Synthesis (HLS) can be used to reduce this lack of flexibility. For example, SystemC Compiler [4] from Synopsys, and Monet from Mentor Graphics, propose a set of I/O scheduling modes (cycle-fixed, superstate, free-floating) that allow to target alternative architectural solutions. Communication is specified using *wait* statements and is mixed with the computation specification what limits the flexibility of the input behavioral description. In these two tools, memory accesses are represented as multi-cycle operations in a Control and Data Flow Graph (CDFG). Memory vertices are scheduled as operative vertices by considering conflicts among data accesses. In practice, the number of nodes in their input specifications must be limited to obtain a realistic and satisfying architectural solution. This limitation is mainly due to the complexity of the algorithms that are used for the scheduling. Only a few works really schedule the memory accesses [5], [6]. They include precise temporal models of those accesses, and try to improve performances without considering the possibility of simultaneous accesses that would ease the subsequent task of register and memory allocation.

In the domain of real-time and data-intensive applications, processing resources have to deal with ever growing data streams. The system/architecture design has therefore to focus on avoiding bottlenecks in the buses and I/O buffers for data-transfer, while reducing the cost of data storage and satisfying strict timing constraints and high-data rates. The methodology that can permit such a design must rely on (1) constraint modeling for both I/O timing and internal data memory, (2) constraint analysis for feasibility checking and (3) high-level synthesis.

In [7] and [8], we proposed a methodology for SoC design that is based on the re-using of algorithmic description. Our approach is based on high-level synthesis techniques under I/O timing constraints and aims to optimally design the corresponding component by taking into account the system integration constraints: the data rate, the technology, and I/O timing properties. In [9], we have introduced a new approach to take into account the memory architecture and the memory mapping in the behavioral synthesis of real-time VLSI circuits. A memory-mapping file was used to include those memory constraints in our HLS tool GAUT [10]. In this paper, we propose a design flow based on formal models that allow high-level synthesis under both I/O timing and memory constraints for digital signal processing algorithms. DSP systems designers specify the I/O timing, the computation latency, the memory distribution and the application's data rate requirements that are the constraints for the synthesis of the hardware components.

This paper is organized as follows: In section 2 we formulate the problem of synthesis under I/O timing and memory constraints. Section 3 presents the main steps of our approach, and its underlying formal models. In section 4, we demonstrate the efficiency of our approach with the didactic example of the Fast Fourier Transform (FFT).

## II. PROBLEM FORMULATION

In this section, we illustrate the inter-dependency between the access parallelism to memory and the timing performances as well as the influence of these two parameters on the resulting component architecture. Let us consider a hardware component based on a generic architecture composed of two main functional units: one memory unit *MU* and one processing unit *PU*. Suppose the computation processed to be $c = (a*v1 + v3)-(b*v2+v4)$ where $v1$, $v2$, $v3$ and $v4$ are variables values stored in memory. Fig. 1(a) shows the *Signal Flow Graph (SFG)* of this algorithm. This component receives input data *a* and *b* from the environment through an input port and sends its result *c* on the output port. All the data used and produced by the processing unit are respectively read and written in a fixed order sequence $S = (a,b,c)$: i.e. $t_a < t_b < t_c$. The read sequence of two variables $v1$ and $v2$ is completely deterministic i.e.: $t_{v1} < t_{v2}$. with $t_{v1} = t_a$ and $t_{v2} = t_b$. However, a scheduling choice is needed to access data v3 and v4 since a single memory bank is available in the component.

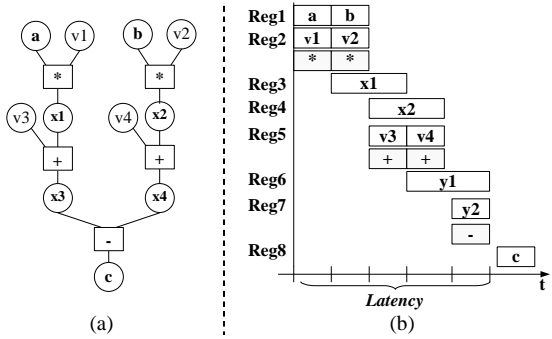

Fig. 1: (a) Signal Flow Graph SFG, (b) Timing behavior,

In our example, we choose to access $v3$ before $v4$. In this context, the minimum latency is therefore equal to 5 cycles (Fig. 1(b)). Fig. 2 presents a possible corresponding architecture of the processing unit that includes 1 multiplier, 1 adder, 1 substractor and 8 registers.

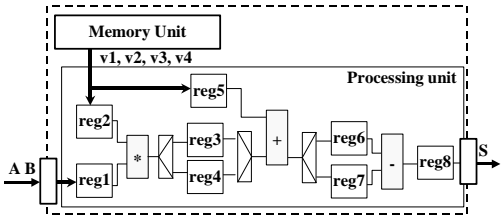

Fig. 2: Sequential architecture

Let us now consider the following data transfer sequence $S_{busses} = (a \,|\, b, c)$: i.e. $t_a = t_b < t_c$. If the latency required to produce the result is long enough ($\geq 5$ cycles) to allow a reordering (serialization) of input data $a$ and $b$, then the previously designed architecture including one memory bank can be reused. However, this solution need to design an input wrapper composed of 1 register, 1 multiplexer and 1 controller. If the required latency is not long enough (i.e. = 3 cycles), the designer must design a new component including 2 multipliers, 2 adders, 11 registers and 2 memory banks (see Fig. 3). In such a case, because of their restricted customization capabilities, neither a pre-designed component nor a macro generator would be flexible enough to respond to the new design constraints.

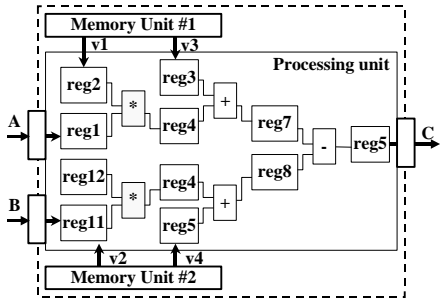

Fig. 3: Parallel architecture

As stated before, a new design flow, based on synthesis under constraints, is needed to get flexibility and to make the DSP component design easier. This includes (1) modeling styles to represent I/O timing and memory constraints, (2) analysis steps to check the feasibility of the constraints (3) methods and techniques for optimal synthesis.

## III. DESIGN APPROACH OVERVIEW

The input of our HLS tool [10] is an algorithmic description that specifies the functionality disregarding implementation details. This initial description is compiled to obtain an intermediate representation: the Signal Flow Graph *SFG* (see Fig. 4).

### A. Timing Constraint Graph

In a first step, we generate an Algorithmic Constraint Graph *ACG* from the operator latencies and the data dependencies expressed in the *SFG*. The latencies of the operators are assigned to operation vertices of the *ACG* during the operator's selection step in the behavioral synthesis flow. Starting from the system description and its architectural model, the integrator, for each bus or port that connects the component to others in the SoC, specifies I/O rates, data sequence orders and transfer timing information. We defined a formal model named *IOCG (IO Constraint Graph)* that supports the expression of integration constraints for each bus (id. port) of the component. Finally we generate a Global Constraint Graph *(GCG)* by merging the *ACG* with the *IOCG* graph. Merging is done by mapping the vertices and associated constraints of *IOCG* onto the input and output vertices set of *ACG*. A minimum timing constraint on output vertices (earliest date for data transfer) of the *IOCG* are transformed into the *GCG* in maximum timing constraints (latest date for data computation/production).

After having described the behavior of the component and the design constraints in a formal model, we analyze the feasibility between the application rate and the data dependencies of the algorithm, in function of the technological constraints. We analyze the I/O timing specifications according to the algorithmic ones: we check if the required constraints on output data are always verified with the behavior specified for input data. The entry point of the IP core design task is the global constraint graph *GCG*.

### B. Memory Constraint Graph

As outlined in the previous subsection, A Signal Flow Graph (*SFG*) is first generated from the algorithmic specification. A Memory Constraint Graph is a cyclic directed polar graph $MCG(V',E',W')$ where $V'=\{v'0,..., v'n\}$ is the set of data vertices placed in memory. A memory Constraint Graph contains $|V'|=n+1$ vertices which represent the memory size, in term of memory elements. The set of edges $E'=(v'i, v'j)$ represents possible consecutive memory accesses, and $W'$ is a function that represents the access delay between two data nodes. $W'$ has only two possible values: *Wseq* (sequential) for an adjacent memory access in memory, or *Wrand* (randomize) for a non adjacent memory access. In our approach, this *SFG* is parsed and a memory table is created. All data vertices are extracted from the *SFG* to construct the memory table. The designer can choose the data to be placed in memory and defines a memory mapping. For every memory in the memory table, we construct a weighted Memory Constraint Graph (*MCG*). It represents conflicts and scheduling possibilities between all nodes placed in this memory. The *MCG* is constructed from the *SFG* and the memory mapping file. It will be used during the scheduling process.

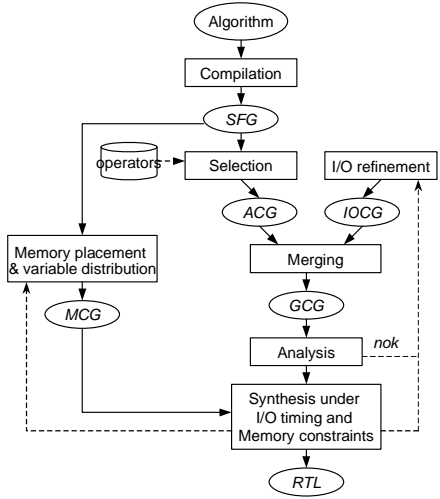

Fig. 4: Proposed Synthesis Flow

Fig. 6(b) shows a MCG for the presented example with one simple port memory bank. The variable data *v1, v2, v3* and *v4* are placed consecutively in one bank. Dotted edges represent *sequential* accesses (two adjacent memory addresses) and plain edges represent random accesses (non-adjacent addresses). Further information about the formal models and the memory design can be found in [7], [8], [9].

### C. Scheduling under I/O and Memory Constraints

The classical "list scheduling" algorithm relies on heuristics in which ready operations (operations to be scheduled) are listed by priority order. In our tool, an early scheduling is performed on the *GCG*. In this scheduling, the priority function depends on the mobility criterion. For operations that have the same mobility, the priority is defined using the operation margin. Next, operations are scheduled and bind to operators (see Fig. 5).

```
Scheduling_Function
1)      Operation_Mobility_computing(GCG)
2)      For (time = 0; time < End; time = time + t_cycle)
3)         List = Operation_Priority_listing(GCG)
4)         Ready_Ops = Find_schedulable_operation(List, time)
5)         Binding(Ready_Ops, operators_set, MCG, time)
6)      End for

Binding Function
1)      While (Ready_Ops!= NULL)
2)        Ops_low_mobility = Get_first(Ready_Ops)
3)        if(Op_low_mobility->margin > 0)
4)           If(Find_mem_conflic(MCG, Ops_low_mobility) = FALSE)
5)              If(operators_set != NULL)
6)                 Ops_Binding(sh_list, operator)
7)              else //no opr or mem conflict
8)                 Posponed(Ops_low_mobility)
9)        else // margin = 0
10)          If(Find_mem_conflict(MCG, Ops_low_mobility) = FALSE)
11)             Operator_cretation()
12)             Ops_Binding(sh_list, operator)
13)          else
14)             Exit(cycle, operator, operation, memory bank, …)
15)          end if
16)     End while
```

Fig. 5: Pseudo code of the scheduling algorithm

An operation can be scheduled if the current cycle is greater than the ASAP time. Whenever two ready operations need to access the same resource (this is a so-called resource conflict), the operation with the lower mobility has the highest priority and is scheduled. The other is postponed. When the mobility is equal to zero, one new operator is allocated to this operation. To perform a scheduling under memory constraint, we introduce memory access operators and add an accessibility criterion based on the *MCG*. A memory has as much access operators as access ports. The list of ready operations is still organised according to the mobility criterion, but all the operations that do not match the accessibility condition are removed from this list. Hence, when the mobility is equal to zero, the synthesis process exits and the designer have to target an alternative solution for the component architecture by reviewing the memory mapping and/or modifying some communication features.

Our scheduling technique is illustrated in Fig. 6 using the previously presented example where the timing constraints are now the following: $S = (a|b,c)$ i.e. $t_a = t_b < t_c$. The memory table (Fig. 6(a)) is extracted from the *SFG* and is used by the designer to define the memory mapping. Internal data *v1, v2, v3* and *v4* are respectively placed at address *@0, @1, @2* and *@3* in the bank0. Our tool constructs one Memory Constraint Graph *MCG* (Fig. 6(b)). In addition to the mapping constraint the designer also specifies two latency *Lat1=5 cycles* and *Lat2=3cycles*.

**For latency Lat1**, the sequential access sequence is $v1 \rightarrow v2 \rightarrow v3 \rightarrow v4$ : it includes 3 dotted edges (with weight Wseq). To deal with the memory bank access conflicts, we define a table of accesses for each port of a memory bank. In our example, the table has only one line for the single memory bank0. The table of memory access has Data_rate / Sequential_access_time elements. The value of each element of the table indicates if a memory access operator is idle or not at the current time (control step c_step). We use the *MCG* to produce a scheduling that permits to access the memory in burst mode. If two operations have the same priority ( margin = Lat1-T(+)-T(*) = 1 cycles) and request the same memory bank, the operation that is scheduled is the operation that involves an access at an address that follows the preceding access. For example, multiplication operation *(a*v1)* and *(b*v2)* have the same mobility. At c_step *cs_1*, they are both executable and the both operands *v1* and *v2* are stored in bank0. *MCG_1* indicates that the sequence $v1 \rightarrow v2$ is shorter than $v2 \rightarrow v1$. We then schedule *(a*v1)* at c_step *cs_1* and *(b*v2)* at c_step *cs_2* to favour the sequential access (see Fig. 6 (c)). At c_step *cs_3*, addition *(x1+v3)* and *(x2+v4)* have the same mobility, the MCG indicates that sequence $v2 \rightarrow v3$ is shorter than $v2 \rightarrow v4$. Addition *(x1+v3)* is scheduled at c_step *cs_3* and *(x2+v4)* at c_step *cs_3*.

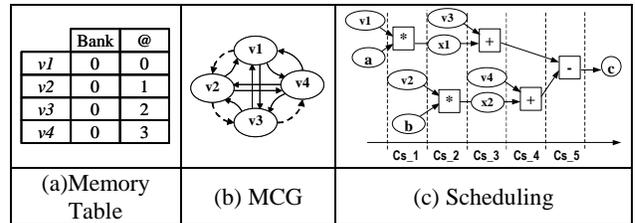

| | Bank | @ |
|---|---|---|
| *v1* | 0 | 0 |
| *v2* | 0 | 1 |
| *v3* | 0 | 2 |
| *v4* | 0 | 3 |

| (a) Memory Table | (b) MCG | (c) Scheduling |
|---|---|---|

Fig. 6: Scheduling under I/O timing and latency constraint

**For latency Lat2,** multiplication operation *(a*v1)* and *(b*v2)* have the same mobility that is null. Both operations must then be scheduled in c_step *cs_1*. Because of the memory access conflict, there is no solution to the scheduling problem: the designer has hence to review its design constraints. He can target an alternative solution by adding one memory bank or by increasing the computing latency.

## IV. EXPERIMENTAL RESULTS

We described in the two previous sections our synthesis design flow and the scheduling under I/O timing and memory constraints. We present now the results of synthesis under constraints obtained using the HLS tool *GAUT* [10]. The algorithm used for this experience is a Fast Fourier Transform (FFT). This FFT reads 128 real input Xr(k) and produces the output Y(k) composed of two parts: one real Yr(k) and one imaginary Yi(k). The *SFG* includes 16897 edges and 8451 vertices. Several syntheses have been realized using a 200MHz clock frequency and a technological library in which the multiplier latency is 2 cycles and the latency of the adder and the subtractor is 1 cycle.

### A. *Experiment 1:* Synthesis under I/O timing constraints

In this first experiment we synthesized the FFT component under I/O timing constraints and analyzed the requirements on memory banks. In order to generate a global constraint graph *GCG*, minimum and maximum timing constraints have been introduced between I/O vertices of the *ACG* graph using the *IOCG* model. The FFT latency is defined by a maximum timing constraint between the first input and the first output vertices. The specified latency (that is the shortest one according to the data dependencies and the operator latencies) corresponds to a 261 cycles delay. The FFT component is constrained to read one *Xr* sample and to produce one Y sample every cycle.

The resulting FFT component contains 20 multipliers, 8 adders and 10 subtractors (see **Exp#1** at Table 1). 8 memory banks are required for those I/O timing constraints. *However, the internal coefficients are mapped in a non-linear scheme in memory. A large amount of memory bank is needed to get enough parallel accesses to reach the specified latency. Moreover, coefficients can possibly be located in multiple banks what requires the design of a complex memory unit.*

### B. *Experiment 2:* Synthesis under memory constraints

In this second experiment we synthesized a FFT component only under memory constraints. Only the maximal number of concurrent access to the memory banks limits the minimal latency. Thus, with a large amount of operators, a latency equal to the critical path delay of the SFG could be obtained. For this reason, we synthesized the FFT with the same number of operators than in the first experiment. Then, we analyzed the requirement on I/O ports and computation latency. The memory constraints are the following: 2 memory banks respecting a simple mapping constraint: the 128 real coefficient Wr in bank0 and the 128 imaginary coefficient Wi in bank1.

The shortest latency imposed by the memory mapping and the number of operators corresponds to a 215 cycles delay (**Exp#2** at Table 1). This delay is shorter that the delay obtained in the previous experiment. This architecture requires 36 input busses and 14 outputs. *However, a large amount of busses with non-trivial data ordering (non-linear data index progression) is needed. If the environment imposes the exchange of data over a smaller number of I/O busses, a communication unit should be designed. This unit would be able to add extra latency to serialize data.*

### C. *Experiment 3:* Synthesis under I/O timing and memory constraints

In this last experiment, we synthesized the FFT component under both I/O timing and memory constraints. We kept the memory mapping used for the second experiment and founded the shortest latency that allows to respect the I/O rates defined in the first experiment. The resulting architecture contains 17 multipliers, 8 adders and 10 subtractors (see **Exp#3** at Table 1). It produces its first result after 343 cycles.

TABLE 1: SYNTHESIS RESULTS

|  | Memory bank. | Input busses | Output busses | Sub. | Add. | Mult. | Latency (in cycle) |
|---|---|---|---|---|---|---|---|
| **Exp#1** | 8 | 1 | 2 | 10 | 8 | 20 | 261 |
| **Exp#2** | 2 | 36 | 14 | 10 | 8 | 20 | 215 |
| **Exp#3** | 2 | 1 | 1 | 10 | 8 | 17 | 343 |

Because of both the memory mapping and the I/O constraints, the latency is greater than in experiment 1 and 2. However, the architecture complexity is equivalent to the previous ones in term of operators. Hence, it appears that synthesis under both I/O timing and memory constraints allows to manage both the system's communication and memory, while keeping a reasonable architecture complexity.

## V. CONCLUSION

In this paper, a design methodology for DSP component under I/O timing and memory constraints is presented. This approach, that relies on constraints modeling, constraints analysis, and synthesis, helps the designer to efficiently implement complex applications. Experimental results in the DSP domain show the interest of the methodology and modeling, that allow tradeoffs between the latency, I/O rate and memory mapping. We are currently working on heuristic rules that could help the designer in exploring more easily different architectural solutions, while considering memory mapping and I/O timing requirements.

## ACKNOWLEDGEMENTS

These works have been realized within the French RNRT Project ALIPTA.